# A lightweight three-user secure quantum summation protocol without a third party based on single-particle states

Tian-Yu Ye*, Tian-Jie Xu

College of Information & Electronic Engineering, Zhejiang Gongshang University, Hangzhou 310018, P.R.China

E-mail：happyyty@aliyun.com(T.Y.Ye)

***Abstract*—In this paper, a lightweight three-user secure quantum summation protocol is put forward by using single-particle states, which can accomplish the goal that three users cooperate together to calculate the modulo 2 addition of their private messages without the help of a third party. This protocol only requires single-particle states rather than quantum entangled states as the initial quantum resource, and only needs single-particle measurements and Bell basis measurements. This protocol needs none of quantum entanglement swapping, the Pauli operations, the controlled-not (CNOT) operation, the Hadamard gate or a pre-shared private key sequence. Security analysis proves that this protocol is secure against both the outside attacks and the participant attacks. Compared with the existing two-dimensional three-user quantum summation protocols, this protocol more or less takes advantage over them on the aspects of the initial quantum resource, users' quantum measurement, the usage of quantum entanglement swapping, the usage of Pauli operations, the usage of CNOT operation or the usage of Hadamard gate.**

## I. INTRODUCTION

It is popularly accepted that secure quantum summation is one of the most fundamental branches of quantum secure computation. For example, it may be used in various scenarios such as quantum anonymous voting [1], quantum private comparison [2], and quantum key agreement [3], and so on. The task of secure quantum summation is to ensure the correctness of summation result and avoid the leakage of private inputs. During recent years, secure quantum summation has naturally gained more and more interests because of its importance. In the year of 2002, Heinrich started the research of quantum summation and applied it into integration [4]. Later, he also investigated quantum Boolean summation with repetitions in the case of worst-average setting [5]. So far, numerous quantum summation protocols [1,6-25] have been constructed from different viewpoints. According to the dimension of quantum system applied, they can be classified into two kinds: two-dimensional quantum summation protocols [6-12] and high dimensional quantum summation protocols [1,12-25]. Among the two-dimensional quantum summation protocols [6-12], some ones have a third party, but others have none. Among the two-dimensional three-user quantum summation protocols [6,8,9,11,12], the protocols in Refs.[6,8,11,12] need to employ quantum entangled states as the initial quantum resource, which are more difficult to prepare than single-particle states in practice; the protocol of Ref.[8] needs to perform the $\left\{\gamma_j^1 \,|\, j=1,2,\ldots,8\right\}$ basis measurements and the $\left\{\gamma_j^2 \,|\, j=1,2,\ldots,8\right\}$ basis measurements, which are much more difficult to realize than the $\left\{|0\rangle,|1\rangle\right\}$ basis measurements and the Bell basis measurements; the protocol of Ref.[6] needs quantum entanglement swapping; the protocols of Refs.[6,11,12] require the Pauli operations; the protocol of Ref.[9] needs the controlled-not (CNOT) operation; and the protocols of Refs.[9,11] require the Hadamard gate. Apparently, as for a quantum communication protocol, if the burdens of the initial quantum resource, the quantum measurement and the auxiliary quantum gates are alleviated, its practical implementation feasibility may be greatly improved.

Based on the above analysis, this paper is devoted to designing a lightweight three-user secure quantum summation protocol without a third party by using single-particle states as the initial quantum resource, where three users can cooperate together to calculate the modulo 2 addition of their private messages by themselves. This protocol more or less has better performance than the previous two-dimensional three-user quantum summation protocols of Refs.[6,8,9,11,12] on the aspects of the initial quantum resource, users' quantum measurement, the usage of quantum entanglement swapping, the usage of Pauli operations, the usage of CNOT operation or the usage of Hadamard gate.

## II. Preliminary Knowledge

The four Bell states can be depicted as

$$\left|\phi^+\right\rangle=\frac{1}{\sqrt{2}}\left(|00\rangle+|11\rangle\right)=\frac{1}{\sqrt{2}}\left(|+\rangle|+\rangle+|-\rangle|-\rangle\right), \tag{1}$$

$$\left|\phi^-\right\rangle=\frac{1}{\sqrt{2}}\left(|00\rangle-|11\rangle\right)=\frac{1}{\sqrt{2}}\left(|+\rangle|-\rangle+|-\rangle|+\rangle\right), \tag{2}$$

$$\left|\psi^+\right\rangle=\frac{1}{\sqrt{2}}\left(|01\rangle+|10\rangle\right)=\frac{1}{\sqrt{2}}\left(|+\rangle|+\rangle-|-\rangle|-\rangle\right), \tag{3}$$

$$\left|\psi^-\right\rangle=\frac{1}{\sqrt{2}}\left(|01\rangle-|10\rangle\right)=\frac{1}{\sqrt{2}}\left(|+\rangle|-\rangle-|-\rangle|+\rangle\right), \tag{4}$$

where $|\pm\rangle=\left(|0\rangle\pm|1\rangle\right)/\sqrt{2}$ . It can be easily derived from Eqs.(1-4) that

$$|0\rangle|0\rangle=\frac{1}{\sqrt{2}}\left(\left|\phi^+\right\rangle+\left|\phi^-\right\rangle\right), \tag{5}$$

$$|0\rangle|1\rangle=\frac{1}{\sqrt{2}}\left(\left|\psi^+\right\rangle+\left|\psi^-\right\rangle\right), \tag{6}$$

$$|1\rangle|0\rangle=\frac{1}{\sqrt{2}}\left(\left|\psi^+\right\rangle-\left|\psi^-\right\rangle\right), \tag{7}$$

$$|1\rangle|1\rangle=\frac{1}{\sqrt{2}}\left(\left|\phi^+\right\rangle-\left|\phi^-\right\rangle\right), \tag{8}$$

$$|+\rangle|+\rangle=\frac{1}{\sqrt{2}}\left(\left|\phi^+\right\rangle+\left|\psi^+\right\rangle\right), \tag{9}$$

$$|+\rangle|-\rangle=\frac{1}{\sqrt{2}}\left(\left|\phi^-\right\rangle+\left|\psi^-\right\rangle\right), \tag{10}$$

$$|-\rangle|+\rangle=\frac{1}{\sqrt{2}}\left(\left|\phi^-\right\rangle-\left|\psi^-\right\rangle\right), \tag{11}$$

$$|-\rangle|-\rangle=\frac{1}{\sqrt{2}}\left(\left|\phi^+\right\rangle-\left|\psi^+\right\rangle\right). \tag{12}$$

## III. The Proposed Three-user Secure Quantum Summation Protocol

Assume that Alice, Bob and Charlie's private messages are $X=\left(x_1,x_2,\ldots,x_n\right)$ , $Y=\left(y_1,y_2,\ldots,y_n\right)$ and $Z=\left(z_1,z_2,\ldots,z_n\right)$ , respectively, where $x_i,y_i,z_i\in\{0,1\}$ , $i=1,2,\ldots,n$ . They wish to calculate the modulo 2 addition of $X$ , $Y$ and $Z$ , i.e., $X\oplus Y\oplus Z=\left(x_1\oplus y_1\oplus z_1,x_2\oplus y_2\oplus z_2,\ldots,x_n\oplus y_n\oplus z_n\right)$ , where $\oplus$ denotes the modulo 2 addition. We construct the following three-user secure quantum summation protocol to finish this task.

**Step 1:** Bob produces a single-particle sequence $S_B=\left\{B_1,B_2,\ldots,B_{8(n+\delta)}\right\}$ , where $B_j$ is randomly in one of the four states $\left\{|0\rangle,|1\rangle,|+\rangle,|-\rangle\right\}$ , $j=1,2,\ldots,8(n+\delta)$ and $\delta$ is a positive parameter. Afterward, Bob generates $\gamma_b$ decoy photons for checking the existence of an outside eavesdropper, each of which is also randomly in one of the four states $\left\{|0\rangle,|1\rangle,|+\rangle,|-\rangle\right\}$ , and randomly inserts them into $S_B$ to form the new sequence $S_B^{'}$ . Here, $\gamma_b$ is a positive integer. Finally, Bob sends $S_B^{'}$ to Alice with the block transmission technology, which was invented by Ref.[26].

In the same time, Charlie prepares a single-particle sequence $S_C=\left\{C_1,C_2,\ldots,C_{8(n+\delta)}\right\}$ , where $C_j$ is also randomly in one of the four states $\left\{|0\rangle,|1\rangle,|+\rangle,|-\rangle\right\}$ and $j=1,2,\ldots,8(n+\delta)$ . Afterward, in order to check the existence of an outside eavesdropper, Charlie produces $\gamma_c$ decoy photons, each of which is also randomly in one of the four states $\left\{|0\rangle,|1\rangle,|+\rangle,|-\rangle\right\}$ , and randomly mixes them with $S_C$ to compose the new sequence $S_C^{'}$ . Here, $\gamma_c$ is a positive integer. Finally, Charlie transmits $S_C^{'}$ to Alice in the manner of block transmission [26].

**Step 2:** Bob works together with Alice to check the existence of an outside eavesdropper as follows: (1) Bob announces Alice the positions and the preparing bases of decoy photons in $S_B^{'}$ ; (2) Alice uses Bob's preparing bases to measure the corresponding decoy photons and informs Bob of her measurement outcomes; (3) Bob checks whether there exists an outside eavesdropper or not by judging the consistency between the initial states of decoy photons and Alice's measurement outcomes. If there exists no outside eavesdropper, the communication will be kept on; otherwise, the communication will be stopped.

In the meanwhile, Charlie also works together with Alice to check the existence of an outside eavesdropper by using the same approach as above.

**Step 3:** Alice makes the $j$ th particle of $S_B$ and the $j$ th particle of $S_C$ compose the $j$ th particle pair, where $j = 1, 2, \dots, 8(n+\delta)$. Bob and Charlie ask Alice to perform the Bell basis measurement on each particle pair. Then, Alice measures each particle pair with the Bell basis. When her measurement result is $\left|\phi^+\right\rangle$ or $\left|\psi^-\right\rangle$, she publishes it directly; and when her measurement result is $\left|\phi^-\right\rangle$ or $\left|\psi^+\right\rangle$, she publishes 'summation'. Afterward, Bob and Charlie use the measurement results of the particle pairs, each of which has two particles originally prepared in the $\{|+\rangle, |-\rangle\}$ basis, to evaluate Alice's honesty as follows: ① when Alice's measurement result is $\left|\phi^+\right\rangle$ or $\left|\psi^-\right\rangle$, they check whether Alice's measurement result satisfies Eqs.(9-12) or not. If Alice is distrustful, they will stop the communication; otherwise, they will continue the communication; ② they check the times that Alice published 'summation'. Note that for the particle pairs, each of which has two particles originally prepared in the $\{|+\rangle, |-\rangle\}$ basis, the times that Alice published 'summation' should be approximately $n+\delta$. If it is abnormally high, they will stop the communication; otherwise, they will continue the communication.

**Step 4:** Among the remaining particle pairs, the number of particle pairs, each of which has two particles originally prepared in the $\{|0\rangle, |1\rangle\}$ basis, is approximately $2(n+\delta)$. For convenience, among these particle pairs, the ones where Alice published 'summation' are named as message particle pairs. Note the number of message particle pairs is approximately $(n+\delta)$. Bob and Charlie check whether the number of message particle pairs is less than $n$ or not. If this is true, they will stop the communication; otherwise, they will continue the communication.

**Step 5:** Bob and Charlie choose the first $n$ message particle pairs to calculate the summation and inform Alice of their positions. Afterward, Alice, Bob and Charlie derive their one-time pad keys $KA = (ka_1, ka_2, \dots, ka_n)$, $KB = (kb_1, kb_2, \dots, kb_n)$ and $KC = (kc_1, kc_2, \dots, kc_n)$ from these $n$ chosen message particle pairs, respectively. When Alice's measurement result on the $i$ th particle pair is $\left|\phi^-\right\rangle$, then $ka_i = 0$; and when it is $\left|\psi^+\right\rangle$, then $ka_i = 1$. When the original prepared state of Bob's particle in the $i$ th particle pair is $|0\rangle$, then $kb_i = 0$; and when it is $|1\rangle$, then $kb_i = 1$. Similarly, when the original prepared state of Charlie's particle in the $i$ th particle pair is $|0\rangle$, then $kc_i = 0$; and when it is $|1\rangle$, then $kc_i = 1$. Here, $i = 1, 2, \dots, n$. Then, Alice, Bob and Charlie obtain $SA = (sa_1, sa_2, \dots, sa_n)$, $SB = (sb_1, sb_2, \dots, sb_n)$ and $SC = (sc_1, sc_2, \dots, sc_n)$ by computing $sa_i = ka_i \oplus x_i$, $sb_i = kb_i \oplus y_i$ and $sc_i = kc_i \oplus z_i$, respectively, $i = 1, 2, \dots, n$. Subsequently, Alice, Bob and Charlie publish $SA$, $SB$ and $SC$, respectively. Finally, each of Alice, Bob and Charlie calculates $sum_i = sa_i \oplus sb_i \oplus sc_i$ for $i = 1, 2, \dots, n$ to get the whole summation result $Sum = (sum_1, sum_2, \dots, sum_n)$.

## IV. Correctness Analysis

In this protocol, Alice, Bob and Charlie utilize the first $n$ message particle pairs to calculate the summation of $X$, $Y$ and $Z$. Obviously, it can be derived that

$$ka_i \oplus kb_i \oplus kc_i = 0, \qquad (13)$$

where $i = 1, 2, \dots, n$. After Alice, Bob and Charlie publish $SA$, $SB$ and $SC$, respectively, each of them calculates

$$\begin{aligned} sum_i &= sa_i \oplus sb_i \oplus sc_i \\ &= (ka_i \oplus x_i) \oplus (kb_i \oplus y_i) \oplus (kc_i \oplus z_i) \\ &= (ka_i \oplus kb_i \oplus kc_i) \oplus (x_i \oplus y_i \oplus z_i). \end{aligned} \qquad (14)$$

According to Eq.(13), it can be obtained from Eq.(14) that

$$sum_i = x_i \oplus y_i \oplus z_i. \qquad (15)$$

It can be concluded that the correctness of summation result of this protocol can be guaranteed.

## V. Security Analysis

① The outside attack

Apparently, the transmission of $S_B^{'}$ from Bob to Alice is dependent from that of $S_C^{'}$ from Charlie to Alice. Without loss of generality, here only analyze the security of the transmission of $S_B^{'}$ against an outside eavesdropper Eve.

(1) The Trojan horse attacks

The Trojan horse attacks mainly refer to the invisible photon eavesdropping attack [27] and the delay-photon Trojan horse attack [28,29]. As the particles of $S_B^{'}$ only undergo the one-way transmission, this protocol can naturally prevent the Trojan horse attacks from Eve.

(2) The measure-resend attack

As Eve is unknown about the positions of decoy photons, in order to try to get Bob's one-time pad key, after intercepting $S_B^{'}$ sent out from Bob, Eve has to measure all of its particles with the $\{|0\rangle,|1\rangle\}$ basis and send the fresh ones in the same states as found to Alice. As for one decoy photon, if it is in the $\{|0\rangle,|1\rangle\}$ basis, Eve's attack will incur no error; and if it is in the $\{|+\rangle,|-\rangle\}$ basis, Eve's attack will be detected with the probability of $\frac{1}{2}$; in conclusion, the probability that Eve's attack can be detected is $\frac{1}{2}\times\frac{1}{2}=\frac{1}{4}$. For $\gamma_b$ decoy photons, Eve's attack will be detected with the probability of $1-\left(\frac{3}{4}\right)^{\gamma_b}$, which will converge to 1 if $\gamma_b$ is big enough.

(3) The intercept-resend attack

As Eve has no access to the positions of decoy photons, in order to try to get Bob's one-time pad key, after intercepting $S_B^{'}$ sent out from Bob, Eve has to substitute all of its particles with the fake ones she generated beforehand randomly in the $\{|0\rangle,|1\rangle\}$ basis and send them to Alice. As for one decoy photon, if it is in the $\{|0\rangle,|1\rangle\}$ basis, Eve's attack will be detected with the probability of $\frac{1}{2}$; and if it is in the $\{|+\rangle,|-\rangle\}$ basis, Eve's attack will be also detected with the probability of $\frac{1}{2}$; in conclusion, the probability that Eve's attack can be detected is $\frac{1}{2}\times\frac{1}{2}+\frac{1}{2}\times\frac{1}{2}=\frac{1}{2}$. For $\gamma_b$ decoy photons, Eve's attack will be detected with the probability of $1-\left(\frac{1}{2}\right)^{\gamma_b}$, which will converge to 1 if $\gamma_b$ is big enough.

(4) The entangle-measure attack

Eve may try to extract some useful information by entangling her auxiliary particle $|\varepsilon\rangle$ with the one in $S_B^{'}$ through a unitary operation $\hat{E}$. As a result, it can be obtained that

$$\hat{E}|0\rangle|\varepsilon\rangle=\alpha_1|0\rangle|\varepsilon_{00}\rangle+\beta_1|1\rangle|\varepsilon_{01}\rangle, \tag{16}$$

$$\hat{E}|1\rangle|\varepsilon\rangle=\beta_2|0\rangle|\varepsilon_{10}\rangle+\alpha_2|1\rangle|\varepsilon_{11}\rangle, \tag{17}$$

where $\varepsilon_{00}$, $\varepsilon_{01}$, $\varepsilon_{10}$ and $\varepsilon_{11}$ are Eve's probe states, and $|\alpha_i|^2+|\beta_i|^2=1$ for $i=1,2$. Consequently, it can be derived that

$$\begin{aligned}\hat{E}|+\rangle|\varepsilon\rangle&=\frac{1}{\sqrt{2}}\left(\alpha_1|0\rangle|\varepsilon_{00}\rangle+\beta_1|1\rangle|\varepsilon_{01}\rangle+\beta_2|0\rangle|\varepsilon_{10}\rangle+\alpha_2|1\rangle|\varepsilon_{11}\rangle\right)\\&=\frac{1}{\sqrt{2}}\left[|0\rangle\left(\alpha_1|\varepsilon_{00}\rangle+\beta_2|\varepsilon_{10}\rangle\right)+|1\rangle\left(\beta_1|\varepsilon_{01}\rangle+\alpha_2|\varepsilon_{11}\rangle\right)\right]\\&=\frac{1}{2}\left[\left(|+\rangle+|-\rangle\right)\left(\alpha_1|\varepsilon_{00}\rangle+\beta_2|\varepsilon_{10}\rangle\right)+\left(|+\rangle-|-\rangle\right)\left(\beta_1|\varepsilon_{01}\rangle+\alpha_2|\varepsilon_{11}\rangle\right)\right]\\&=\frac{1}{2}\left[|+\rangle\left(\alpha_1|\varepsilon_{00}\rangle+\beta_2|\varepsilon_{10}\rangle+\beta_1|\varepsilon_{01}\rangle+\alpha_2|\varepsilon_{11}\rangle\right)+|-\rangle\left(\alpha_1|\varepsilon_{00}\rangle+\beta_2|\varepsilon_{10}\rangle-\beta_1|\varepsilon_{01}\rangle-\alpha_2|\varepsilon_{11}\rangle\right)\right],\end{aligned} \tag{18}$$

$$\begin{aligned}\hat{E}|-\rangle|\varepsilon\rangle&=\frac{1}{\sqrt{2}}\left(\alpha_1|0\rangle|\varepsilon_{00}\rangle+\beta_1|1\rangle|\varepsilon_{01}\rangle-\beta_2|0\rangle|\varepsilon_{10}\rangle-\alpha_2|1\rangle|\varepsilon_{11}\rangle\right)\\&=\frac{1}{\sqrt{2}}\left[|0\rangle\left(\alpha_1|\varepsilon_{00}\rangle-\beta_2|\varepsilon_{10}\rangle\right)+|1\rangle\left(\beta_1|\varepsilon_{01}\rangle-\alpha_2|\varepsilon_{11}\rangle\right)\right]\\&=\frac{1}{2}\left[\left(|+\rangle+|-\rangle\right)\left(\alpha_1|\varepsilon_{00}\rangle-\beta_2|\varepsilon_{10}\rangle\right)+\left(|+\rangle-|-\rangle\right)\left(\beta_1|\varepsilon_{01}\rangle-\alpha_2|\varepsilon_{11}\rangle\right)\right]\\&=\frac{1}{2}\left[|+\rangle\left(\alpha_1|\varepsilon_{00}\rangle-\beta_2|\varepsilon_{10}\rangle+\beta_1|\varepsilon_{01}\rangle-\alpha_2|\varepsilon_{11}\rangle\right)+|-\rangle\left(\alpha_1|\varepsilon_{00}\rangle-\beta_2|\varepsilon_{10}\rangle-\beta_1|\varepsilon_{01}\rangle+\alpha_2|\varepsilon_{11}\rangle\right)\right].\end{aligned} \tag{19}$$

In order for not being detected by Alice and Bob, according to Eqs.(16-19), it should satisfy that

$$\beta_1=\beta_2=0, \tag{20}$$

$$\alpha_1|\varepsilon_{00}\rangle+\beta_2|\varepsilon_{10}\rangle-\beta_1|\varepsilon_{01}\rangle-\alpha_2|\varepsilon_{11}\rangle=0, \tag{21}$$

$$\alpha_1|\varepsilon_{00}\rangle-\beta_2|\varepsilon_{10}\rangle+\beta_1|\varepsilon_{01}\rangle-\alpha_2|\varepsilon_{11}\rangle=0. \tag{22}$$

It can be deduced from Eqs.(20-22) that

$$\alpha_1|\varepsilon_{00}\rangle=\alpha_2|\varepsilon_{11}\rangle. \tag{23}$$

Consequently, in order for not being detected by Alice and Bob, the final state of Eve's auxiliary particle should be independent from the particle of $S_B^{'}$, which makes Eve acquire nothing by launching this kind of attack.

② The participant attack

The participant attack, which is generally more powerful than the outside attack [30], is further analyzed here. Naturally, in a quantum summation protocol with $p$ users, when any $p-1$ users conspire together, they can easily derive the left one's privacy from the final summation result. Here, $p$ is a positive integer equal or greater than 3. Hence, it is only necessary to analyze the participant attack from one dishonest user.

(1) The participant attack from Alice

Alice can hear $SB$ and $SC$ from Bob and Charlie, respectively. Apparently, in order to decode out $Y$ and $Z$, Alice should know $KB$ and $KC$ in advance, respectively. Hence, Alice may adopt the following attack strategy: in Step 3, she employs the $\{|0\rangle,|1\rangle\}$ basis rather than the Bell basis to measure two particles in each of the remaining $8(n+\delta)$ particle pairs, and publishes Bob and Charlie the fake Bell basis measurement result in hope of escaping the honesty check against her. For one particle pair for checking Alice's honesty, when her measurement result is $|0\rangle|0\rangle$ or $|1\rangle|1\rangle$, she randomly publishes the fake measurement result $|\phi^+\rangle$ or 'summation'; and when her measurement result is $|0\rangle|1\rangle$ or $|1\rangle|0\rangle$, she randomly publishes the fake measurement result $|\psi^-\rangle$ or 'summation'. Without loss of generality, suppose that the particle pair is $|+\rangle|+\rangle$. When Alice's measurement result is $|0\rangle|0\rangle$ or $|1\rangle|1\rangle$, according to Eqs.(9-12), Alice will be detected with the probability of 0 if she publishes the fake measurement result $|\phi^+\rangle$, and will be also detected with the probability of 0 if she publishes 'summation', as there is no check for this case; when Alice's measurement result is $|0\rangle|1\rangle$ or $|1\rangle|0\rangle$, according to Eqs.(9-12), Alice will be detected with the probability of 1 if she publishes the fake measurement result $|\psi^-\rangle$, and will be detected with the probability of 0 if she publishes 'summation', as there is no check for this case. To sum up, when the particle pair is $|+\rangle|+\rangle$, the probability that Alice can be detected is $\frac{1}{4}\times\frac{1}{2}\times1+\frac{1}{4}\times\frac{1}{2}\times1=\frac{1}{4}$. For $2(n+\delta)$ particle pairs for checking Alice's honesty, the probability that Alice can be detected is $1-\left(\frac{3}{4}\right)^{2(n+\delta)}$, which will converge to 1 if $2(n+\delta)$ is large enough.

Moreover, Alice may also adopt the following attack strategy: in Step 3, she employs the $\{|0\rangle,|1\rangle\}$ basis rather than the Bell basis to measure two particles in each of the remaining $8(n+\delta)$ particle pairs, and always publishes 'summation' in hope of escaping the honesty check against her. As a result, the times that Alice published 'summation' for the particle pairs, each of which has two particles originally prepared in the $\{|+\rangle,|-\rangle\}$ basis, is approximately $2(n+\delta)$. When Bob and Charlie check the times that Alice published 'summation' for the particle pairs, each of which has two particles originally prepared in the $\{|+\rangle,|-\rangle\}$ basis, they will easily discover Alice's this kind of cheating behavior.

(2) The participant attack from Bob or Charlie

In this protocol, the role of Bob is identical to that of Charlie. Without loss of generality, Bob is supposed to be dishonest. Bob can hear $SA$ and $SC$ from Alice and Charlie, respectively. Apparently, in order to decode out $X$ and $Z$, Bob should know $KA$ and $KC$ in advance, respectively. $KA$ and $KC$ are produced from the message particle pairs. As a result, in order to obtain $KC$, Bob may launch his attack as follows: he intercepts $S_C^{'}$ sent out from Charlie, uses the $\{|0\rangle,|1\rangle\}$ basis to measure the particles in $S_C^{'}$ whose counterparts in $S_B^{'}$ were prepared by him in the $\{|0\rangle,|1\rangle\}$ basis, and sends the fresh ones in the same states as found together with other ones in $S_C^{'}$ to Alice. However, Bob inevitably leaves his trace on the decoy photons in $S_C^{'}$. Concretely speaking, as for one decoy photon, the probability that Eve's attack can be detected is $\frac{1}{2}\times\frac{1}{2}=\frac{1}{4}$; hence, Eve's attack will be detected with the probability of $1-\left(\frac{3}{4}\right)^{\tau}$, where $\tau$ is the number of decoy photons measured by Bob.

In addition, Bob has no opportunity to deduce $KA$ from $KB$ and $KC$, since he cannot get $KC$ beforehand without being detected. Furthermore, Bob has no chance to know $KA$ just by Alice's publishment of 'summation'.

## VI. Discussions

We further compare this protocol with the two-dimensional three-user quantum summation protocols in Refs.[6,8,9,11,12] where three users participate in the calculation of modulo 2 addition of their private messages. The comparison results are listed in Table 1 for clarity. Note that we ignore the quantum resource and the classical resource consumed for the security check against Eve in Table 1, as it often can be regarded to be relatively independent from the main part of a quantum communication protocol.

In Table 1, the qubit efficiency is expressed as [29]

$$\lambda=\frac{v}{q+r}, \qquad (24)$$

where $v$ represents the number of classical bits to be calculated for summation, $q$ denotes the number of qubits consumed, and $r$ represents the number of classical bits consumed.

In this protocol, each of $X$, $Y$ and $Z$ has $n$ bits, hence $v=n$; both $S_B$ and $S_C$ have $8(n+\delta)$ single particles randomly in the four states $\{|0\rangle,|1\rangle,|+\rangle,|-\rangle\}$, so $q=8(n+\delta)\times 2=16(n+\delta)$; and Alice, Bob and Charlie need to publish $SA$, $SB$ and $SC$, respectively, hence $r=3n$. Consequently, the qubit efficiency of this protocol is $\lambda=\frac{n}{16(n+\delta)+3n}$.

In the protocol of Ref.[6], the secret bit string from the user $A_i$ has $L$ bits, where $i=1,2,3$, hence $v=L$; $A_1$ needs to produce $L$ three-particle entangled states, while both $A_2$ and $A_3$ need to generate $L$ single photons, so $q=5L$; both $A_2$ and $A_3$ need to cooperate to publish the classical bits coded from their Bell basis measurement results, while $A_1$ needs to publish the final summation result, in order to let $A_2$ and $A_3$ know it, hence $r=3L$. Consequently, the qubit efficiency of the protocol of Ref.[6] is $\lambda=\frac{L}{5L+3L}=\frac{1}{8}$.

In the protocol of Ref.[8], the secret bit string from the user $P_i$ has $L$ bits, where $i=1,2,3$, hence $v=L$; $P_1$ needs to prepare $\left\lceil \frac{L}{2} \right\rceil+\delta$ genuinely maximally six-particle entangled states, so $q=6\left(\left\lceil \frac{L}{2} \right\rceil+\delta\right)$; and $P_i$ needs to publish $C_i$, hence $r=3L$. Consequently, the qubit efficiency of the protocol of Ref.[8] is $\lambda=\frac{L}{6\left(\left\lceil \frac{L}{2} \right\rceil+\delta\right)+3L}$.

In the protocol of Ref.[9], the user $P_i$ has one secret bit, where $i=1,2,3$, hence $v=1$; $P_i$ needs to produce three single particle sequences, each of whose length is $1+d$, so $q=3(1+d)$; and $P_i$ needs to publish the ciphertext $c_i$, hence $r=3$. Consequently, the qubit efficiency of the protocol of Ref.[9] is $\lambda=\frac{1}{3(1+d)+3}$.

In the protocol of Ref.[11], the secret bit string from the user $P_i$ has $L$ bits, where $i=1,2,3$, hence $v=L$; $P_1$ needs to prepare $L$ three-particle entangled states, so $q=3L$; and in order to make the other two users know the final summation result, $P_1$ needs to publish it, hence $r=L$. Consequently, the qubit efficiency of the protocol of Ref.[11] is $\lambda=\frac{L}{3L+L}=\frac{1}{4}$.

In the protocol of Ref.[12], the secret bit string from the user $A_i$ has $m$ bits, where $i=1,2,3$, hence $v=m$; each user needs to generate $m$ Bell states, so $q=6m$; and $A_1$ needs to publish the result $P_i$, hence $r=3m$. Consequently, the qubit efficiency of the protocol of Ref.[12] is $\lambda=\frac{m}{6m+3m}=\frac{1}{9}$.

According to Table 1, we can draw the following conclusions: ① As for the initial quantum resource, this protocol exceeds the protocols of Refs.[6,8,11,12], as the preparations of Bell states and multiple-particle entangled states are much more difficult than that of single-particle states; ② as for users' quantum measurement, this protocol takes advantage over the protocol of Ref.[8], as the $\left\{\gamma_j^1 \middle| j=1,2,\ldots,8\right\}$ basis measurements and the $\left\{\gamma_j^2 \middle| j=1,2,\ldots,8\right\}$ basis measurements are much more complicated to realize than the $\{|0\rangle,|1\rangle\}$ basis measurements and the Bell basis measurements; ③ as for the usage of quantum entanglement swapping, this protocol defeats the protocol of Ref.[6], as it doesn't need to employ quantum entanglement swapping; ④ as for the usage of Pauli operations, this protocol has better performance than the protocols of Refs.[6,11,12], as it doesn't need to adopt the Pauli operations; ⑤ as for the usage of CNOT operation, this protocol exceeds the protocol of Ref.[9], as it doesn't require the CNOT operation; ⑥ as for the usage of Hadamard gate, this protocol exceeds the protocols of Refs.[9,11], as it doesn't need the Hadamard gate.

Moreover, the comparison results of the proposed protocol and the semiquantum summation protocols of Refs.[32,33] are further summarized in Table 2.

## VII. CONCLUSIONS

In conclusion, a lightweight three-user secure quantum summation protocol without the existence of a third party based on single-particle states is proposed in this paper, which can calculate the modulo 2 addition of three users' private messages. Single-particle states rather than quantum entangled states are employed as the initial quantum resource. Three users are only required to implement single-particle measurements and Bell basis measurements. None of quantum entanglement swapping, the Pauli operations, the CNOT operation or the Hadamard gate are required. Moreover, three users are not required to share a private key sequence among them beforehand. This protocol is secure against both the outside attacks and the participant attacks. Compared with the previous two-dimensional three-user quantum summation protocols of Refs.[6,8,9,11,12], this protocol more or less takes advantage over them on the aspects of the initial quantum resource, users' quantum measurement, the usage of quantum entanglement swapping, the usage of Pauli operations, the usage of CNOT operation or the usage of Hadamard gate.

Besides, the proposed protocol needs to perform the complete Bell basis measurement. At present, it is hard to realize in linear optics but can be achieved via cross-Kerr nonlinearities or cavity-assisted light-matter interactions [34-36].

Moreover, the proposed protocol can be extended into the collective-dephasing noise resistant version and the collective-rotation noise resistant version, as long as the single-particle sequences generated in Step 1 are composed by the logical qubits $\{|0_{dp}\rangle, |1_{dp}\rangle, |+_{dp}\rangle, |-_{dp}\rangle\}$ and $\{|0_r\rangle, |1_r\rangle, |+_r\rangle, |-_r\rangle\}$, respectively. Here, $|0_{dp}\rangle = |01\rangle$, $|1_{dp}\rangle = |10\rangle$, $|+_{dp}\rangle = \frac{1}{\sqrt{2}}(|0_{dp}\rangle + |1_{dp}\rangle)$, $|-_{dp}\rangle = \frac{1}{\sqrt{2}}(|0_{dp}\rangle - |1_{dp}\rangle)$, $|0_r\rangle = |\phi^+\rangle$, $|1_r\rangle = |\psi^-\rangle$, $|+_r\rangle = \frac{1}{\sqrt{2}}(|0_r\rangle + |1_r\rangle)$ and $|-_r\rangle = \frac{1}{\sqrt{2}}(|0_r\rangle - |1_r\rangle)$ [37-40].

In addition, semiquantum summation [32,33,41], which is the semiquantum counterpart of quantum summation, has been put forward recently. In the future, it is also worthy of studying how to design secure semiquantum summation protocols.

## Acknowledgment

The authors would like to thank the anonymous reviewers for their valuable comments that help enhancing the quality of this paper. Funding by the National Natural Science Foundation of China (Grant No.62071430 and No.61871347), the Fundamental Research Funds for the Provincial Universities of Zhejiang (Grant No.JRK21002) and Zhejiang Gongshang University, Zhejiang Provincial Key Laboratory of New Network Standards and Technologies (No. 2013E10012) is gratefully acknowledged.

TABLE 1 COMPARISON RESULTS OF DIFFERENT TWO-DIMENSIONAL THREE-USER QUANTUM SUMMATION PROTOCOLS

| | The protocol of Ref.[6] | The protocol of Ref.[8] | The protocol of Ref.[9] | The protocol of Ref.[11] | The protocol of Ref.[12] | This protocol |
|---|---|---|---|---|---|---|
| The existence of a third party | No | No | No | No | No | No |
| Initial quantum resource | Single-particle states and GHZ states | Six-particle entangled states | Single-particle states | Three-particle entangled states | Bell states | Single-particle states |
| Users' quantum measurement | $\{\lvert+\rangle,\lvert-\rangle\}$ basis measurements and Bell basis measurements | $\{\lvert 0\rangle,\lvert 1\rangle\}$ basis measurements, $\{\lvert+\rangle,\lvert-\rangle\}$ basis measurements, Bell basis measurements, $\{\gamma_j^1 \mid j=1,2,\ldots,8\}$ basis measurements and $\{\gamma_j^2 \mid j=1,2,\ldots,8\}$ basis measurements | $\{\lvert 0\rangle,\lvert 1\rangle\}$ basis measurements and $\{\lvert+\rangle,\lvert-\rangle\}$ basis measurements | $\{\lvert 0\rangle,\lvert 1\rangle\}$ basis measurements | $\{\lvert 0\rangle,\lvert 1\rangle\}$ basis measurements | $\{\lvert 0\rangle,\lvert 1\rangle\}$ basis measurements and Bell basis measurements |
| Usage of quantum entanglement swapping | Yes | No | No | No | No | No |
| Usage of Pauli operations | Yes | No | No | Yes | Yes | No |
| Usage of CNOT operation | No | No | Yes | No | No | No |
| Usage of Hadamard gate | No | No | Yes | Yes | No | No |
| Usage of a pre-shared private key sequence | No | No | No | No | No | No |
| Summation type | Modulo 2 addition | Modulo 2 addition | Modulo 2 addition | Modulo 2 addition | Modulo 2 addition | Modulo 2 addition |
| Computation way | Bit-by-bit | Bit-by-bit | Bit-by-bit | Bit-by-bit | Bit-by-bit | Bit-by-bit |
| Qubit efficiency | $\frac{1}{8}$ | $\frac{L}{6\left(\left\lceil \frac{L}{2} \right\rceil + \delta\right) + 3L}$ | $\frac{1}{3(1+d)+3}$ | $\frac{1}{4}$ | $\frac{1}{9}$ | $\frac{n}{16(n+\delta)+3n}$ |

TABLE 2 COMPARISON RESULTS OF THE PROPOSED PROTOCOL AND THE PROTOCOLS OF REFS.[32,33]

| | The protocol of Ref.[32] | The protocol of Ref.[33] | This protocol |
|---|---|---|---|
| Quantum channel | ideal quantum channel | ideal quantum channel | ideal quantum channel |
| Function | calculating the summation of the private bits from three classical participants | calculating the summation of the private bits from one quantum participant and two classical participants | calculating the summation of the private bits from three quantum participants |
| Number of involved participants | four | three | three |
| Initial quantum resource | $\lvert+\rangle$ | $\lvert+\rangle$ | $\{\lvert 0\rangle,\lvert 1\rangle,\lvert+\rangle,\lvert-\rangle\}$ |
| Quantum measurement of quantum participant | $\{\lvert 0\rangle,\lvert 1\rangle\}$ basis measurements, $\{\lvert+\rangle,\lvert-\rangle\}$ basis measurements and GHZ-type basis measurements | $\{\lvert 0\rangle,\lvert 1\rangle\}$ basis measurements, $\{\lvert+\rangle,\lvert-\rangle\}$ basis measurements and Bell basis measurements | $\{\lvert 0\rangle,\lvert 1\rangle\}$ basis measurements, $\{\lvert+\rangle,\lvert-\rangle\}$ basis measurements and Bell basis measurements |
| Quantum measurement of classical participants | $\{\lvert 0\rangle,\lvert 1\rangle\}$ basis measurements | $\{\lvert 0\rangle,\lvert 1\rangle\}$ basis measurements | / |
| Usage of quantum entanglement swapping | No | No | No |
| Usage of unitary operations | No | No | No |
| Usage of a pre-shared private key | No | No | No |
| Summation type | modulo 2 addition | modulo 2 addition | modulo 2 addition |
| Computation way | bit-by-bit | bit-by-bit | bit-by-bit |